\documentclass[aps,amsmath,amssymb,prl,reprint,superscriptaddress]{revtex4-1}

\usepackage{bm,graphicx}

\begin{document}

\newcommand{\rev}[1] {\textcolor{green}{#1}}
\newcommand{\erl}[1] {\textcolor{blue}{#1}}
\newcommand{\mre}[1] {\textcolor{red}{#1}}

\newcommand{\mgb}{MgB$_2$}
\newcommand{\Tc}{T_{\text{c}}}
\newcommand{\Hcii}{H_{\text{c2}}}
\newcommand{\Hac}{H_{\text{AC}}}
\newcommand{\Hdc}{H_{\text{DC}}}
\newcommand{\degrees}{^{\circ}}

\newcommand{\fms}{f_{\text{MS}}}
\newcommand{\Ims}{I_{\text{MS}}}
\newcommand{\Iesi}{I_{\text{ES}_1}}
\newcommand{\Iesii}{I_{\text{ES}_2}}
\newcommand{\fes}{f_{\text{ES}}}

\title{Structural Transition Kinetics and Activated Behavior in the Superconducting Vortex Lattice}

\author{E.~R.~Louden}
\author{C.~Rastovski} 
\affiliation{Department of Physics, University of Notre Dame, Notre Dame, Indiana 46656, USA}

\author{S.~J.~Kuhn}
\affiliation{Department of Physics, University of Notre Dame, Notre Dame, Indiana 46656, USA}
\affiliation{Center for Exploration of Energy and Matter, Indiana University, Bloomington, Indiana 47408, USA}

\author{A.~W.~D.~Leishman}
\affiliation{Department of Physics, Kent State University, Kent, Ohio 44240, USA}
\affiliation{Department of Physics, University of Notre Dame, Notre Dame, Indiana 46656, USA}

\author{L.~DeBeer-Schmitt}
\affiliation{Oak Ridge National Laboratory, Oak Ridge, Tennessee 37831, USA}

\author{C.~D.~Dewhurst}
\affiliation{Institut Laue-Langevin, 71 avenue des Martyrs, CS 20156, F-38042 Grenoble cedex 9, France}

\author{N.~D.~Zhigadlo}
\affiliation{Laboratory for Solid State Physics, ETH, CH-8093 Zurich, Switzerland}
\affiliation{Department of Chemistry and Biochemistry, University of Bern, CH-3012 Bern, Switzerland}

\author{M.~R.~Eskildsen}
\altaffiliation{Corresponding author: eskildsen@nd.edu}
\affiliation{Department of Physics, University of Notre Dame, Notre Dame, Indiana 46656, USA}

\date{\today}

\begin{abstract}
Using small-angle neutron scattering, we investigated the behavior of a metastable vortex lattice state in {\mgb} as it is driven towards equilibrium by an AC magnetic field.
This shows an activated behavior, where the AC field amplitude and cycle count are equivalent to, respectively, an effective ``temperature'' and ``time''.
The activation barrier increases as the metastable state is suppressed, corresponding to an aging of the vortex lattice.
Furthermore, we find a cross-over from a partial to a complete suppression of metastable domains depending on the AC field amplitude, which may empirically be described by a single free parameter.
This represents a novel kind of collective vortex behavior, most likely governed by the nucleation and growth of equilibrium vortex lattice domains.
\end{abstract}

\maketitle

Structural phase transformations are ubiquitous in solids, exhibiting common features independent of the microscopic properties of a particular material.
It is natural to expect that this commonality extends to vortex matter in type-II superconductors.
Vortices are of great fundamental interest~\cite{Huebener:2001we,Blatter:1994gz,Giamarchi:1995tq}, exhibiting similarities with skyrmions~\cite{Pfleiderer_SkL_rev,Nagaosa:2013cc};
soft matter systems such as liquid crystals, colloids and granular materials~\cite{Nagel:2017fe}; and glasses~\cite{Lubchenko:2007cc}.
Each vortex carries one quantum of magnetic flux, and vortex matter
thus constitutes an ideal, mono-disperse system.
As such, it presents a conceptually simple two-dimensional model system to examine fundamental problems such as structure formation and transformation at the mesoscopic scale,
metastable states, and non-equilibrium kinetics.

The presence of metastable non-equilibrium vortex lattice (VL) phases in superconducting {\mgb} is well established~\cite{Das:2012cf}.
The metastability is not due to vortex pinning, but represents a novel kind of collective vortex behavior, most likely due to the presence of VL domain boundaries~\cite{Rastovski:2013ff}.
Similar field/temperature history dependent metastability has also recently been reported in connection with skyrmion lattice transitions~\cite{Makino:2017hh,Nakajima:2017uc,Bannenberg:2017ws}.
More broadly, domain nucleation and growth governs the behavior of a wide range of physical systems,
and one may expect similarities between the VL and e.g. martensitic phase transitions~\cite{Wang:2017jw} or domain switching in ferroelectrics~\cite{Shin:2007gu}.
However, the kinetics of structural VL phase transformations has remained largely unexplored both experimentally and theoretically~\cite{Deutsch:2010bi,Durkin:2017aa}.

\begin{figure*}
    \includegraphics[width = 0.8 \textwidth]{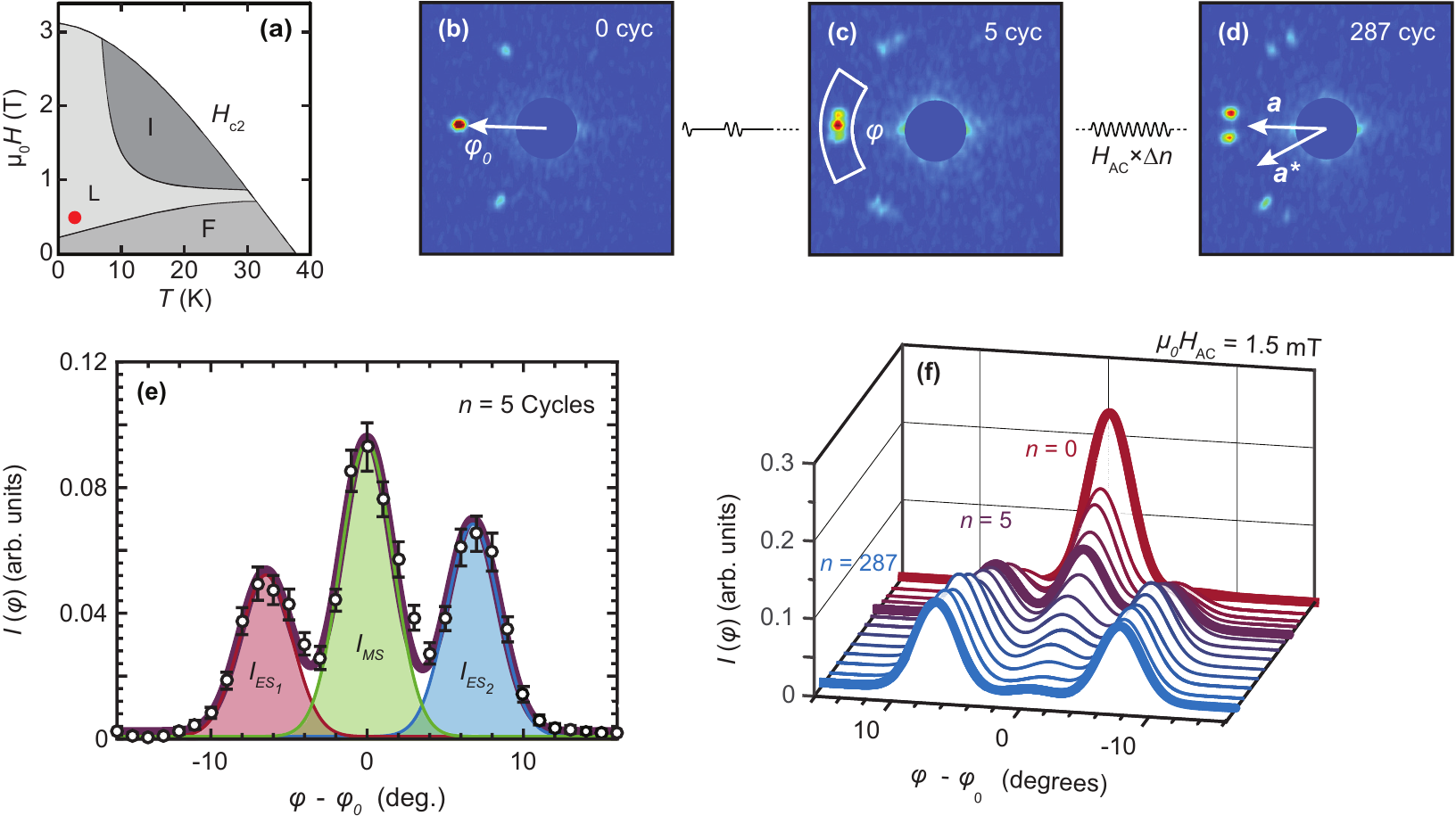}
    \caption{
        Vortex lattice evolution from the MS to the ES with $\mu_0 \Hac = 1.5$~mT.
        (a) Equilibrium VL phase diagram.
        Diffraction patterns show the pristine MS F phase (b), an intermediate state after application of 5~AC cycles (c), and the ES L phase obtained after 287 cycles (d).
        Each shows the same region of reciprocal space, with the VL reference direction ($\varphi_0$) and crystalline axes indicated in (b) and (d) respectively.
        Background scattering near the center of the detector is masked off.
        (e) Azimuthal intensity distribution, comprised of neutron counts from the detector area indicated in (c), fitted with a three-peak Gaussian.
        (f) Fits for the entire MS to ES measurement sequence, with curves corresponding to data in (b-d) highlighted in bold.
        Each curve is separately normalized to its total intensity.
        }
    \label{Fig1}
\end{figure*}

Here we report the results of small-angle neutron scattering (SANS) measurements,
using AC magnetic fields of varying amplitude to gradually drive the VL from the metastable state (MS) to the equilibrium state (ES).
The equilibrium VL phase diagram for the hexagonal superconductor {\mgb}, shown in Fig.~\ref{Fig1}(a), is composed of three hexagonal configurations denoted F, L and I~\cite{Das:2012cf,Hirano:2013jx}.
In both the F [Fig.~\ref{Fig1}(b)] and I phases a single global orientational order is observed in the VL diffraction patterns,
indicated by six Bragg peaks aligned with respectively the $\bm{a}$ 
and $\bm{a^*}$ direction within the crystalline basal plane.
In the intermediate L phase [Fig.~\ref{Fig1}(d)], the VL rotates continuously from the $\bm{a}$ to the $\bm{a^*}$ orientation, where the presence of both clockwise and counterclockwise domain rotations leads to 12 Bragg peaks. 
Cooling or heating across the F-L or L-I phase transitions leaves the VL in a metastable state, as thermal excitations are insufficient to drive the system to equilibrium~\cite{Das:2012cf}.

Measurements were performed on the GP-SANS beam line at the High Flux Isotope Reactor at Oak Ridge National Laboratory, and the D33 beam line at Institut Laue-Langevin.
The final data presented here was collected at D33~\cite{5-42-420}, but consistent results were found at both facilities.
The SANS measurements were performed in a conventional VL setup, with both the incident neutrons and the magnetic field parallel to the neutron beam~\cite{MuhlbauerRMP}.
We used the same 200~$\mu$g single crystal of {\mgb} ($\Tc = 38$~K, $\mu_0 \Hcii = 3.1$~T) as in prior studies~\cite{Karpinski:2003fe,Das:2012cf, Rastovski:2013ff}.
However, VL metastability was confirmed in other crystals of {\mgb}~\cite{Das:2012cf}, and recently also in samples from a different source~\cite{ManniXtal}.
Further details concerning the SANS experiment are given in the Supplemental Material~\cite{SM}.

A typical SANS measurement sequence is illustrated in Fig.~\ref{Fig1}, including representative VL diffraction patterns obtained at various stages along the transition.
All measurements were performed at 2.6~K with 0.5~T applied parallel to the crystal $\bm{c}$ axis. 
Prior to each measurement sequence, a pristine MS VL was prepared as follows.
First, an equilibrium VL was obtained in the F phase ($T > 13.2$~K), 
by performing a damped oscillation of the DC magnetic field with an initial amplitude of 50~mT~\cite{Das:2012cf}.
As the VL density is directly proportional to the applied field, this leads to a breathing motion where vortices are pushed into and out of the sample.
In superconductors with low pinning, this results in a well-ordered, equilibrium VL configuration~\cite{Levett:2002ba}.
In the present case of {\mgb}, this is confirmed by the sharp Bragg peaks evident in Fig.~\ref{Fig1}(b)-(d) and discussed in more detail in the Supplemental Material~\cite{SM}.
Following the damped field oscillation the ES VL was cooled to 2.6~K across the F-L phase boundary to obtain a MS.

We used a stop-motion technique,
alternating between imaging the VL by SANS and using a bespoke coil to apply a controlled number ($n$) of 250~Hz AC field cycles with amplitudes  ($\mu_0 \Hac$) between 0.5 and 1.5~mT parallel to the DC field used to create the VL.
The small AC field, roughly two orders of magnitude less than the damped oscillation used to prepare the pristine MS,
causes the VL to evolve gradually towards the ES and allows for a detailed study of the relaxation process.
The pristine MS VL [Fig.~\ref{Fig1}(b)], imaged before the application of any AC cycles, shows a single F phase Bragg peak along the $\varphi_0$ direction.
After application of 5 AC cycles [Fig.~\ref{Fig1}(c)], the coexistence of both MS and ES ($\varphi - \varphi_0 \approx \pm 7^{\circ}$) Bragg peaks are observed.  
Following a total of 287 cycles [Fig.~\ref{Fig1}(d)], only peaks corresponding to the equilibrium L phase are seen clearly.
To make efficient use of the neutron beam time measurements were performed at a single angular setting, with only peaks at the left side of the detector satisfying the Bragg condition.

For a quantitative analysis, we consider the azimuthal intensity distribution $I(\varphi)$, fitted by a three-peak Gaussian as shown in Fig.~\ref{Fig1}(e)~\cite{SM}.
Here each peak corresponds to the scattering contribution from one of the three possible VL domain orientations.
Fits for the entire MS to ES measurement sequence are shown in Fig.~\ref{Fig1}(f).
As the number of applied AC cycles is increased, the central Bragg peak associated with the MS F phase gradually loses intensity, and outer ES L phase peaks emerge.
At the end of the measurement sequence, the L phase is achieved within a majority of the sample, with only a small remnant intensity at $\varphi = \varphi_0$.
We note that the MS to ES transition proceeds in a discontinuous manner, with VL domains nucleating at one of the two equilibrium orientations and subsequently growing at the expense of the metastable domains.
This is in striking contrast to the continuous transition observed for the equilibrium VL, where domains gradually rotate away from $\varphi = \varphi_0$ as a function of magnetic field and/or temperature~\cite{Das:2012cf}.
    
\begin{figure}
    \centering
    \includegraphics{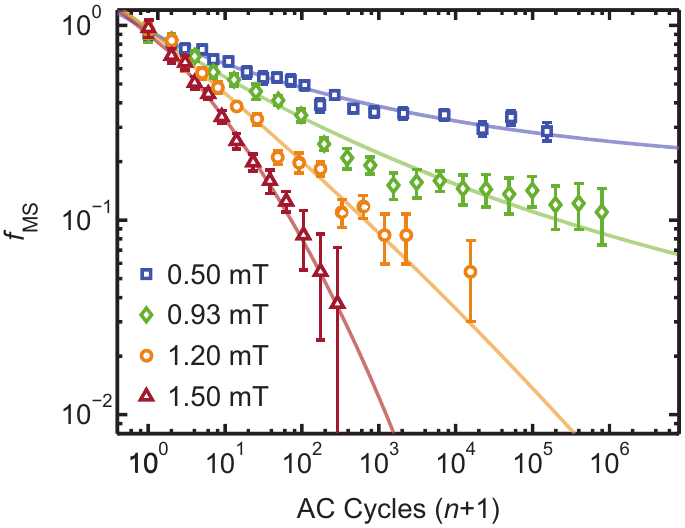}
    \caption{
    Metastable VL volume fraction as a function of applied cycles for different AC field amplitudes.
    The AC cycle count is offset by one to allow $\fms(0)$ for the pristine VL to be included.
    Lines are fits to Eq.~(\ref{Eq:fMSfit}).}
    \label{Fig2}
\end{figure}

The integrated intensity of the Bragg peaks obtained from the fits, is proportional to the number of vortices within each of the corresponding VL domain orientations.
Thus, the volume fraction of the sample remaining in the MS can be determined by dividing the intensity of the F phase ($\varphi = \varphi_0$) peak with the total scattered intensity:
\begin{equation}
    \fms = \frac{\Ims}{\Iesi+\Ims+\Iesii}.
\end{equation}
Figure~\ref{Fig2} shows the metastable volume fraction versus the number of applied cycles for the different AC field amplitudes used in this work.
With increasing $\Hac$, fewer AC cycles are required to suppress $\fms$.
Moreover, the curvature of $\fms(n)$, plotted on a log-log scale, changes from positive ($0.50, 0.93$~mT) to negative ($1.50$~mT).

\begin{figure}
    \centering
    \includegraphics{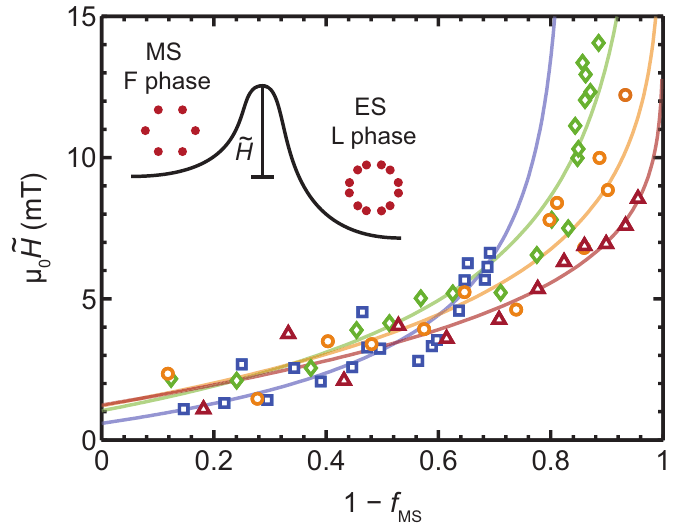}
    \caption{
    Activation field determined using Eq.~(\ref{Eq:act}) and the data in Fig.~\ref{Fig2}; the symbols correspond to the same values of $\Hac$ as previously.
    Lines are calculated using Eqs.~(\ref{Eq:Htilde}) and (\ref{Eq:Htilde0}), and the parameters obtained from the fits of $\fms(n)$ in Fig.~\ref{Fig2}.
    The inset illustrates the activation barrier between the metastable and equilibrium VL states.}
    \label{Fig3}
\end{figure}

The presence of long lived metastable phases in {\mgb} cannot be understood based on the single domain free energy~\cite{Das:2012cf}.
Rather, the presence of additional energy barriers are required to prevent the individual VL domains from rotating to the equilibrium orientation, illustrated in the inset to Fig.~\ref{Fig3}.
The absence of any thermally driven relaxation towards the ES within experimental time scales is consistent with the small Ginzburg number $Gi \sim 10^{-6}$ for {\mgb}~\cite{Klein:2010hh}.
This also implies that the properties of the initial MS VL prepared before each measurement sequence will not depend on the F phase temperature or the cooling rate.
It is also important to note that transient behavior of the MS to ES transition sets these results apart from conventional vortex dynamics focusing on the current carrying capabilities of superconductors,
with the latter focusing on (non-equilibrium) steady states~\cite{Blatter:1994gz}.

The vortices move in response to the AC field, and it is natural to analyze the SANS data using an activated behavior.
In this context the AC amplitude and cycle count take the role of, respectively, the effective ``temperature" and ``time".
Thus, the decay rate of the metastable volume fraction is given by
\begin{equation}
  \frac{d \fms}{dn} = - \fms \; \exp \left[ -\tilde{H}/\Hac \right],
  \label{Eq:act}
\end{equation}
where $\tilde{H}$ is the activation field representing the barrier between metastable and equilibrium VL domain orientations. 
The proportionality to $\fms$ accounts for the amount of metastable VL volume available for ES domain nucleation and/or growth.

Figure~\ref{Fig3} shows the activation field obtained directly from the data in Fig.~\ref{Fig2}.
Each value of $\tilde{H}$ was determined from two adjacent values of $\fms(n)$ by
$\tilde{H} = - \Hac \ln \left[ - \tfrac{\ln \{\fms(n_{i+1})/\fms(n_i)\}}{n_{i+1} - n_i} \right]$.
Within the scatter of the data, $\tilde{H}$ collapses onto a single curve for $1 - \fms \leq 0.7$,
suggesting a near universal behavior consistent with an activated transition.
Surprisingly, the activation field increases as the metastable volume fraction decreases.
The change of $\tilde{H}$ with ``time'' ($n$), is equivalent to an aging of the VL~\cite{Henkel:2010wn}.
For $1 - \fms \geq 0.7$ the activation field for the different values of $\Hac$ begins to diverge.

To parameterize the MS to ES transition, the metastable volume fraction was fitted by
\begin{equation}
  \fms(n) = \exp \left[ \alpha + \beta (n+1)^{\gamma} \right].
  \label{Eq:fMSfit}
\end{equation}
This functional form is motivated by the saturation of $\fms$ observed for $\mu_0\Hac = 0.50$ and $0.93$~mT. 
For $0 < \gamma < 1$, it corresponds to a stretched exponential function frequently used to model relaxation phenomena~\cite{Klafter:1986gk}.
As seen in Fig.~\ref{Fig2}, the fits provide an excellent description of the data for all AC field amplitudes.
From these, the activation field curves shown in Fig.~\ref{Fig3} may be directly calculated by
\begin{equation}
  \tilde{H} = \tilde{H}_0 - \Hac \left( \frac{\gamma - 1}{\gamma} \right) \; \ln \left[ 1 - \frac{\ln \fms}{\ln f_{\alpha}} \right],
  \label{Eq:Htilde}
\end{equation}
where
\begin{equation}
  \tilde{H}_0 = -\Hac \left[\ln (-\beta \gamma) + \frac{\gamma - 1}{\gamma} \; \ln ( - \alpha/\beta ) \right]
  \label{Eq:Htilde0}
\end{equation}
and $f_{\alpha} = e^{\alpha}$.
When $\alpha$ and $\gamma$ are both negative, $f_{\alpha}$ is the saturation value of $\fms$ as $n \rightarrow \infty$.
The metastable VL states are due to the non-zero value of $\tilde{H}_0$, which prevents a spontaneous onset of the MS to ES transition.

\begin{figure}
    \centering
    \includegraphics{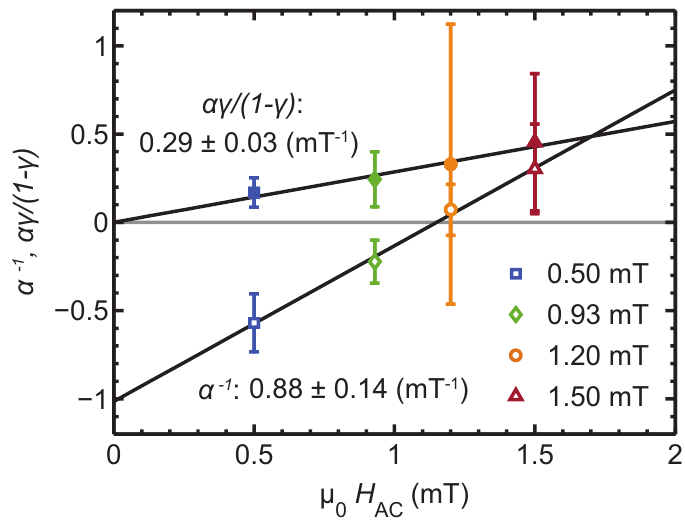}
    \caption{
    Relationship between both $\alpha^{-1}$ and $\alpha \gamma / (1-\gamma)$ and the AC field amplitude.
    Lines are linear fits with slopes indicated in the plot.
    }
    \label{Fig4}
\end{figure}

The cross-over from positive to negative curvature in Fig.~\ref{Fig2} is directly reflected in the evolution of $\alpha^{-1}$, as shown in Fig.~\ref{Fig4}.
Here we find a linear dependence on the AC amplitude and a sign change, yielding a cross-over field $\mu_0 \Hac^* =$ 1.2 $\pm$ 0.3~mT.
In addition, $\alpha \gamma / (1-\gamma)$ (also shown) is found to be directly proportional to the AC field amplitude.
Finally, since $\fms(0) \lesssim 1$ for all values of $\Hac$, $\alpha -\beta \lesssim 0$ for all fits~\cite{SM}.
Consequently, the MS to ES transition is determined by  a single parameter, which depends only on $\Hac$.
Further theoretical studies of VL kinetics, including the effect of domain boundaries, are required to explain this empirical result.
One possible cause for the observed cross-over is a change in the relative magnitude of the activation barriers for the nucleation and growth of the ES VL domains~\cite{Liu:2004dp}.

Due to the many similarities between vortices and skyrmions,
the kinetic properties reported here for the VL may also occur for metastable skyrmion lattice phases~\cite{Makino:2017hh,Nakajima:2017uc,Bannenberg:2017ws}.
Of particular relevance is the possibility of inducing rotational transitions in the skyrmion lattice, either thermally~\cite{Pollath:2017fz} or by the application of AC magnetic fields~\cite{White:2014ji}.
We also note the resemblance of the {\mgb} VL to structural martensitic phase transitions such as the tetragonal-to-orthorhombic transition in cuprate superconductors~\cite{Beyers:1987vq,Chen:1987cn} and the $\alpha$-to-$\epsilon$ transition in iron~\cite{Kalantar:2005hd,Wang:2017jw}.
In these systems, there are two equal energy pathways from the initial to the final structure.
The final configuration is a periodic twin-boundary lattice rather than the lowest energy configuration of a single global domain, with the interface between the initial phase and the twinned phases providing the necessary stabilizing force~\cite{Barsch:1987kk}.
An increasing activation energy, such as the one reported here for the VL, has also been found for isothermal martensitic phase transformations in maraging steel~\cite{Martin:2010mse}.

Finally, glasses are the quintessential example of a supercooled, metastable configuration observed in conjunction with a thermally driven transition.
The metastable VL states in {\mgb} show many parallels to supercooled liquids and other structural glasses, such as an activated transition between states resulting from a complicated energy landscape and a behavior that is governed by domains/domain walls~\cite{Lubchenko:2007cc,Nussinov:2017fm}.
The slowing kinetics (aging) and the diverging activation barrier in Fig.~\ref{Fig3} provides further support for this analogy,
which could be explored further by measuring flux creep in response to transport currents below the vortex de-pinning threshold.

In summary, we have studied the metastable to equilibrium state transition kinetics for the vortex lattice in {\mgb} under the influence of an AC magnetic field.
The kinetics display an activated behavior, with the AC field amplitude and cycle count acting as respectively an effective ``temperature'' and ``time''.
The activation barrier increases gradually as the metastable state is suppressed, indicating an aging of the vortex lattice. 
The transition towards the ES shows a cross-over between partial and complete suppression of metastable domains as the AC field amplitude is increased.
Empirically, we find that the MS to ES transition can be described by a single parameter which is determined by the AC amplitude.

Further studies to provide real space information about VL dislocations, grain boundaries and domain formation,
either experimentally (e.g. by STM) or by non-equilibrium molecular dynamics simulations~\cite{Pollath:2017fz,Olszewski:2018fp},
would be a valuable complement to our SANS results and interpretation.
This would also allow us to connect our observations to results in other classes of materials,
such as the recent study of grain boundary kinetics in two-dimensional triangular colloidal crystals~\cite{Lavergne:2018ds}.

We are grateful to J.~Karpinski for providing the {\mgb} single crystal used for this work.
We acknowledge useful discussions with G.~Blatter, B.~Janko, K.~Newman, Z. Nussinov, M.~Pleimlimg, and U.~C.~T\"{a}uber, and assistance with the SANS experiments from J.~Archer.
This work was supported by the U.S. Department of Energy, Office of Basic Energy Sciences, under Award No. DE-SC0005051.
A portion of this research used resources at the High Flux Isotope Reactor, a DOE Office of Science User Facility operated by the Oak Ridge National Laboratory.

%

\end{document}


\newcommand{\mgb}{MgB$_2$}
\newcommand{\Tc}{T_{\text{c}}}
\newcommand{\Hcii}{H_{\text{c2}}}
\newcommand{\Hac}{H_{\text{AC}}}
\newcommand{\Hdc}{H_{\text{DC}}}
\newcommand{\degrees}{^{\circ}}
\newcommand{\fms}{f_{\text{MS}}}
\newcommand{\Ims}{I_{\text{MS}}}
\newcommand{\Iesi}{I_{\text{ES}_1}}
\newcommand{\Iesii}{I_{\text{ES}_2}}
\newcommand{\fes}{f_{\text{ES}}}
\renewcommand{\thefigure}{SM\arabic{figure}}

\title{Supplementary Material: Structural Transition Kinetics and Activated Behavior in the Superconducting Vortex Lattice}

\author{E.~R.~Louden}
\author{C.~Rastovski} 
\affiliation{Department of Physics, University of Notre Dame, Notre Dame, Indiana 46656, USA}

\author{S.~J.~Kuhn}
\affiliation{Department of Physics, University of Notre Dame, Notre Dame, Indiana 46656, USA}
\affiliation{Center for Exploration of Energy and Matter, Indiana University, Bloomington, Indiana 47408, USA}

\author{A.~W.~D.~Leishman}
\affiliation{Department of Physics, Kent State University, Kent, Ohio 44240, USA}
\affiliation{Department of Physics, University of Notre Dame, Notre Dame, Indiana 46656, USA}

\author{L.~DeBeer-Schmitt}
\affiliation{Oak Ridge National Laboratory, Oak Ridge, Tennessee 37831, USA}

\author{C.~D.~Dewhurst}
\affiliation{Institut Laue-Langevin, 71 avenue des Martyrs, CS 20156, F-38042 Grenoble cedex 9, France}

\author{N.~D.~Zhigadlo}
\affiliation{Laboratory for Solid State Physics, ETH, CH-8093 Zurich, Switzerland}
\affiliation{Department of Chemistry and Biochemistry, University of Bern, CH-3012 Bern, Switzerland}

\author{M.~R.~Eskildsen}
\affiliation{Department of Physics, University of Notre Dame, Notre Dame, Indiana 46656, USA}

\maketitle

In the following we present further information about the experimental details, the azimuthal fitting algorithm and the parameterization of the metastable state (MS) to equilibrium state (ES) vortex lattice (VL) transition.

\subsection*{Experimental details}
The small-angle neutron scattering (SANS) experiment used incoming neutrons, with wavelength $\lambda = 0.7$~nm and wavelength spread $\Delta \lambda/\lambda = 10\%$,
parallel to the DC magnetic field applied by a high-field cryo-magnet.
The beam collimation was defined by a a 10~mm source aperture and a 2~mm sample aperture (slightly larger than the sample size) separated by 12.8~m.
The azimuthal resolution was estimated from the width ($\Delta q$) of the undiffracted beam on the detector.  
Considering a peak of this size at the expected scattering vector $q_{\text{VL}} = 0.105$~nm$^{-1}$ for a vortex lattice at 0.5~T~\cite{MuhlbauerRMP},
the angular azimuthal resolution ($w_{\text{res}}$) is given by
\begin{equation}
	\tan \left( \frac{w_{\text{res}}}{2} \right) = \frac{\Delta q}{2q_{\text{VL}}}.
	\label{res}
\end{equation}
The tight collimation yields a small $w_{\text{res}} = 3.1^{\circ}$~FWHM, required to resolve the closely spaced MS and ES VL Bragg reflections~\cite{Das:2012cf, Rastovski:2013ff}.

During the SANS measurement sequences, vortex motion was induced using a bespoke coil to apply a controlled number of AC field cycles parallel to the DC field used to create the VL.
A sinusoidal wave function was used, with a peak-to-peak amplitude between 0.5 and 1.5~mT and a frequency of 250~Hz.
The low frequency is equivalent to applying a ``fast DC'' field oscillation but with greater precision than can be obtained with a high-field cryo-magnet,
and at the same time achieving a controlled relaxation of the VL affecting the entire sample volume in a reasonable amount of time.

The {\mgb} single crystal used for the SANS experiment was grown using a high pressure cubic anvil technique that has been shown to produce good quality single crystals~\cite{Karpinski:2003fe},
and isotopically enriched $^{11}$B was used to decrease neutron absorption.
The crystal had a flat plate morphology, with an area of $\sim 1 \times 1$~mm$^2$ and a thickness of $\sim 75~\mu$m.
Demagnetization effects are negligible for the used field and measurement geometry as demonstrated by the equality of the applied field ($\mu_0 H$) and measured internal magnetic induction ($B$)~\cite{Rastovski:2013ff}.

\subsection*{Azimuthal fitting algorithm}
To analyze the neutron scattering data, the scattered intensity was binned along the azimuthal direction.
The same detector area, shown in Fig.~1(c) of the main text, was used for the analysis of all measurement sequences.
The azimuthal intensity distribution is well described by a sum of three Gaussians, corresponding to the Bragg peak for each of the VL domain orientations:
\begin{equation}
  	I (\varphi) = I_0 + \sum\limits_{j = 1}^{3} \frac{I_j}{w_j} \; \exp \left[ - 2 \sqrt{\log \, 4} \left( \frac{\varphi - \varphi_{j}}{w_j} \right)^2 \right].
	\label{eq1}
\end{equation}
Here $I_0$ is a constant accounting for isotropic background scattering,  $I_j$ is the integrated intensity, $w_j$ is the full width half maximum (FWHM), and  $\varphi_{j}$ is the center for the $j$th Bragg peak. 

\begin{figure*}[b]
    \centering
    \includegraphics{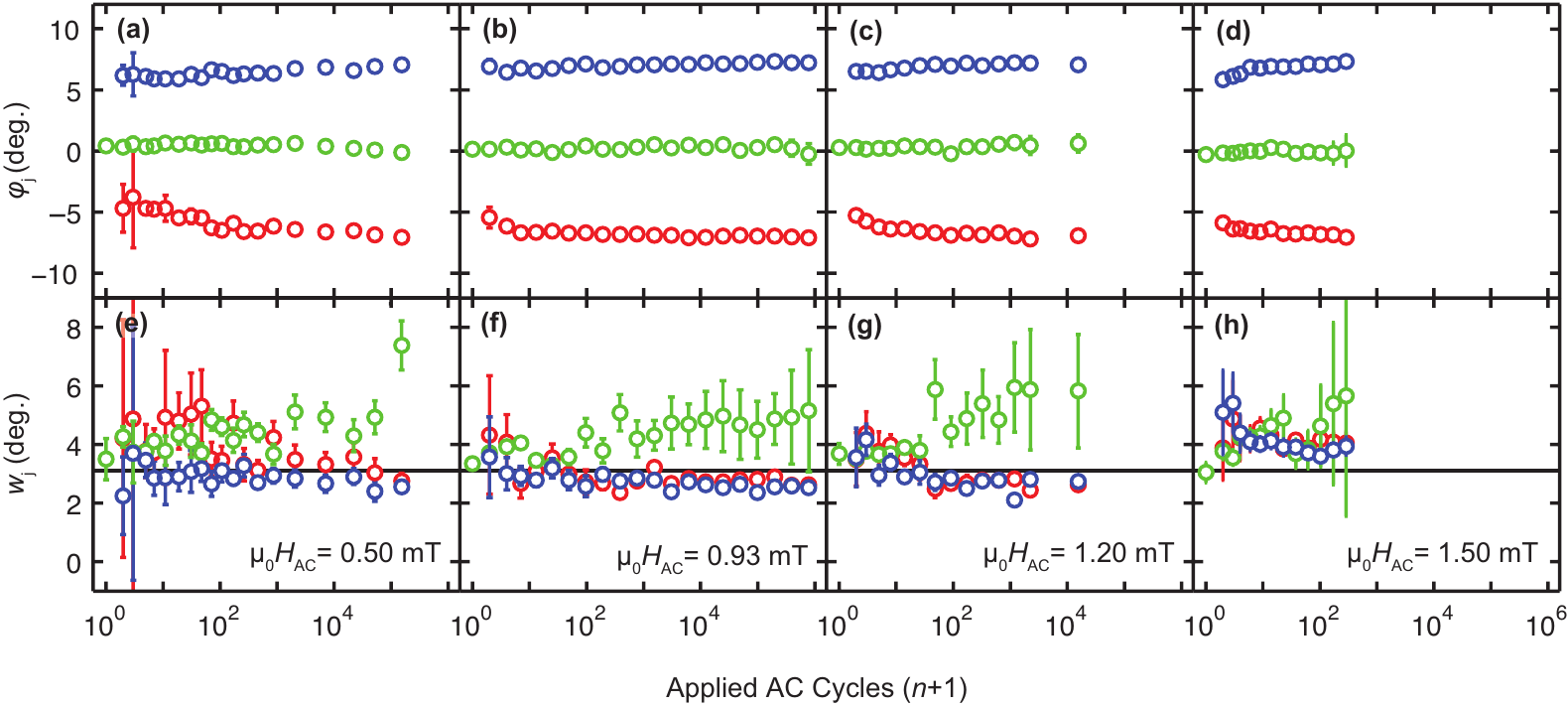}
    \caption{
        Fitted Bragg peak centers (a)-(d) and widths (e)-(h) throughout the MS to ES transition.
        Fits were performed using Eq.~(\ref{eq1}) with no parameter constraints.
        The angular resolution ($w_{\text{res}} = 3.1^{\circ}$) is shown by the black line and colors indicate the different Bragg peaks.
        The number of AC cycles were offset by one in order to display the pristine MS ($n = 0$).
	\label{FigS1}}
\end{figure*}

\begin{figure*}
    \centering
    \includegraphics{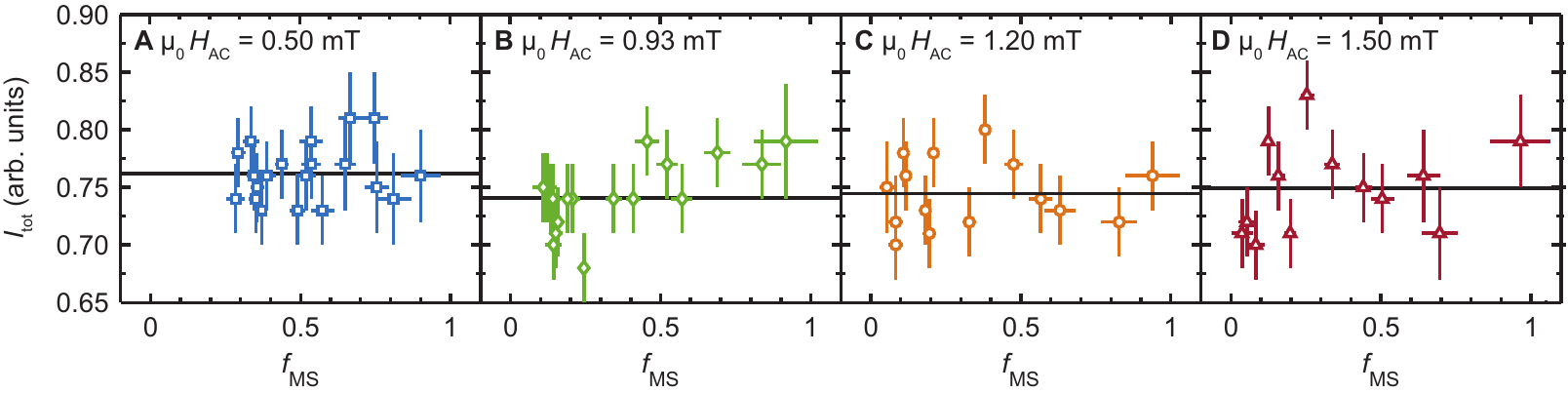}
    \caption{Total measured intensity for the combined MS and ES VL Bragg peaks for different AC field amplitudes. The average values are indicated by the black lines.}
    \label{FigS2}
\end{figure*}

Figure~\ref{FigS1} shows the centers (a)-(d) and widths (e)-(h) obtained from fits to Eq.~(\ref{eq1}).
Once the Bragg peaks for the ES domains are well developed ($n \geq 10^1$) their fitted positions are near constant, and the widths are essentially resolution limited and consistently smaller than those of the metastable domains.
While more ordered ES domains (narrower peaks) would not be surprising, it is not possible to make a definitive conclusion within the precision of the fits.
In particular, we note that error on the fitted peaks centers and widths are greater for the low intensity peaks, such as the ES domains at small $n$ or the MS domains after $\sim 10^2$~AC cycles.
In summary, the data in Fig.~\ref{FigS1} does not indicate that the AC cycles cause a significant disordering of the VL.
This conclusion is supported by Fig.~\ref{FigS2}, which shows the combined scattered intensity of both MS and ES VL Bragg peaks. This remains constant throughout the transition.

Based on the result discussed above, we analyzed the SANS data using a single constant width ($w_j \equiv w$) for all the VL Bragg peaks throughout a given measurement sequence.
While this eliminates potential information contained in the peak widths, this constraint yields more precise fits of the peak positions and intensities.
The following algorithm was used to determine the most appropriate width to use for a given sequence:
first, initial estimates for the peak positions and widths were determined from the clearly resolved peaks in the first  ($n = 0$) and last ($n = n_{\text{max}}$) measurements.
These values were then used to seed independent fits of the individual SANS measurements in the sequence, using a common width for all the peaks.
For the final fit of the data the width was fixed to the average value, calculated from peaks where $I_j$ exceeds 10\% of the combined scattered intensity. 
This average width was $\simeq 4^{\circ}$ FWHM), slightly larger than the estimated instrumental resolution.

\subsection*{Parameterization of MS to ES VL transition}
\begin{figure}
    \centering
    \includegraphics{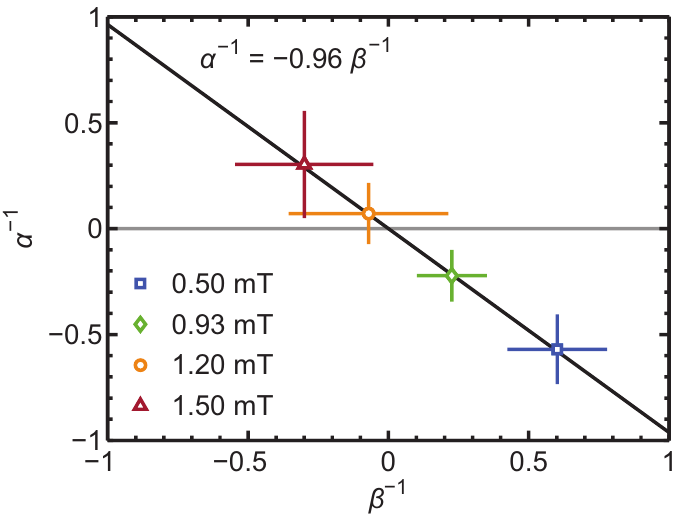}
    \caption{Relationship between $\alpha$ and $\beta$.
            The line is a linear fit to the data, with the slope given in the plot.}
    \label{FigS3}
\end{figure}

The metastable VL volume fraction ($\fms$) was fitted by the expression:
$$
  \fms(n) = \exp \left[ \alpha + \beta (n+1)^{\gamma} \right].
$$
For all the measurements sequences in Fig.~2 of the main text the pristine metastable volume fraction $\fms(0)$ is slightly less than one,
although the deviation from unity is within the error on the determination of $\fms$.
The introduction of separate parameters $\alpha$ and $\beta$ allows us to accommodate this in our fits, although we still expect $\alpha \approx - \beta$.
This is confirmed by Fig.~\ref{FigS3} showing a direct proportionality between $\alpha$ and $\beta$ with a slope close to one.


%